\documentclass[aps,pra,10pt,superscriptaddress,twocolumn]{revtex4-2}
\usepackage{graphicx}
\usepackage{amsmath,amsthm,amssymb,dsfont}
\usepackage{mdframed}
\usepackage{orcidlink}
\usepackage{appendix}
\usepackage{xcolor}
\usepackage[subpreambles=true]{standalone}
\usepackage{maths-common}
\usepackage{complex-numbers}
\usepackage{hilbert-space}
\usepackage{groups}
\let\tensor\undefined 
\usepackage{lie-algebras}
\newcommand{\ket}[1]{\left| #1 \right\rangle}

\newcommand{\ketbra}[2]{\left|#1\right\rangle\hskip-1mm\left\langle#2\right|}

\usepackage{hyperref}
\usepackage[capitalise,nameinlink]{cleveref}
\usepackage[authormarkuptext=name,commentmarkup=uwave]{changes}
\definechangesauthor[name={JRH},color=green!70!black]{jonte}

\begin{document}

\title{How to unitarily map between any two pure states with a single closed-form exponential}
\author{Peter T. J. Bradshaw\,\orcidlink{0000-0001-9938-8460}}
\email{peter.bradshaw@newcastle.ac.uk}
\affiliation{Quantum Group, School of Computing, Newcastle University, 1 Science Square, Newcastle upon Tyne, NE4 5TG, UK}

\author{Marcus Gouveia}
\affiliation{Quantum Group, School of Computing, Newcastle University, 1 Science Square, Newcastle upon Tyne, NE4 5TG, UK}

\author{Jonte R. Hance\,\orcidlink{0000-0001-8587-7618}}
\email{jonte.hance@newcastle.ac.uk}
\affiliation{Quantum Group, School of Computing, Newcastle University, 1 Science Square, Newcastle upon Tyne, NE4 5TG, UK}
\affiliation{Quantum Engineering Technology Laboratories, Department of Electrical and Electronic Engineering, University of Bristol, Woodland Road, Bristol, BS8 1US, UK}

\begin{abstract}
    It is well-known that any two pure quantum states (in the same Hilbert space) can be mapped to any other using unitary transformations. However, previous approaches to this problem required two explicit bases for the Hilbert space, one each for the initial and target states, and thus their complexity necessarily scales with the dimension of the Hilbert space. In this Letter, we show how to utilize novel algebraic methods to construct a closed-form exponential unitary transformation which achieves this in general, using only a single unitary generator. This construction is independent of any bases and agnostic to the dimension of the Hilbert space. We highlight the usefulness of this tool for studying relationships between systems of pure states in quantum information theory, as well in elementary analyses of quantum circuits and unitary operators.
\end{abstract}

\maketitle

\section{Introduction}

It is well-known in the literature on quantum information that any pure state can be mapped to any other pure state living in the same dimension of Hilbert space using a unitary operator, which effectively acts as a higher-dimensional ``rotation'' through the space. However, knowing that a unitary exists which will take you from the state you have to the state you want, and knowing what that unitary is, are two very different things. 

Approaches to find such unitaries typically rely on building one with brute force using the Gram-Schmidt process. This involves constructing two orthonormal bases (one containing the state you have, and another containing the set you want), and then defining a pairwise mapping between the elements of the two basis sets. Concretely, for initial state $\ket{\psi_0}$ and final state $\ket{\phi_0}$, such a process involves using the Gram-Schmidt procedure to find orthonormal bases $\{\psi_i\}$ and $\{\phi_i\}$ for $0\leq i\leq d-1$, where $d$ is the Hilbert space dimension, then defining $\hat{U} = \sum_i \ketbra{\psi_i}{\phi_i}$. However, this approach is messy (especially for large-dimensional systems). It also requires already having a set of linearly independent vectors spanning the space lying around. Finally, and perhaps most egregiously, this brute-force Gram-Schmidt-based approach doesn't appeal to the intrinsic notion of unitary transformations being rotations through Hilbert space - a notion typically manifesting in the unitary being able to be written as the exponential of some Hermitian operator times $i\theta$, where $\theta$ is some angle of rotation.

In this Letter, we instead use algebraic methods to create a different process allowing us to always find the unitary between two pure states. This algebraically-inspired process has two advantages over the Gram-Schmidt-based approach. First, the unitary generated in this way can always be written as a closed-form exponential. Second, this process means we don't first need to use the Gram-Schmidt procedure to define two complete orthogonal basis sets containing the initial and final state. We will begin in section \ref{sec:hilbert} by considering Hilbert space from an abstract perspective; in section \ref{sec:gen}, we will consider the generators of unitary transformations and their elementary realization in Hilbert space; in section \ref{sec:min-pol}, we will derive the properties of a simple unitary generator from its minimal polynomial, and apply these results to solve the present problem in section \ref{sec:exp}; discussion about the advantages and applications of our results will be given in section \ref{sec:disc}. Throughout, our work will be presented independent of basis and dimension-agnostic to maximize the applicability of our results.

\section{Abstract Hilbert Space}\label{sec:hilbert}

Quantum mechanics is written in the mathematical language of Hilbert spaces \cite{teschl_mathematical_2014}. Abstractly, a complex Hilbert space $\generichilbertspace$ is an ordered pair $\generichilbertspacedefinition$ consisting of a vector space $\genericspace$ with underlying set $\genericspaceset$ over $\genericspacefield$, and a positive-definite, conjugate-symmetric, sesquilinear form $\innerproductdeclaration$, i.e. $\forall a,b,c,d\in\genericspaceset, \alpha,\beta\in\genericspacefield$,
\begin{subequations}
\begin{gather}
    a\neq0 \Rightarrow\innerproduct[a][a] > 0 \\
    \innerproduct[b][a] = \complexconjugate{\innerproduct[a][b]} \\
    \begin{aligned}
        \innerproduct[\alpha a + c][\beta b + d] = \complexconjugate{\alpha}&\beta\innerproduct[a][b]+\complexconjugate{\alpha}\innerproduct[a][d] \\&+ \beta\innerproduct[c][b] + \innerproduct[c][d].
    \end{aligned}
\end{gather}
\end{subequations}
The Hilbert space is also typically required to be complete with respect to the inner product, but we will not need to use this property. This abstract structure facilitates the definition of familiar concepts in a more general context, independently of any choice of basis or the dimension of the Hilbert space. In this notation, the Cauchy-Schwartz inequality \cite{teschl_mathematical_2014} reads,
\begin{equation}\label{eqn:cauchy-schwartz}
    \innerproduct[a][a]\innerproduct[b][b]\geq\innerproduct[a][b]\innerproduct[b][a],
\end{equation}
with equality iff $b=\alpha a$ for some $\alpha\in\complexnumbers$; and the Hermitian adjoint $\hermitianadjoint{A}$ of a linear map $A$ is defined as,
\begin{equation}\label{eqn:hermitian-adjoint}
    \innerproduct[A(a)][b]=\innerproduct[a][\hermitianadjoint{A}(b)],
\end{equation}
for all $a,b\in\generichilbertspace$.

Most importantly for this work, we may define both unitary transformations and their associated Lie algebra generators. A unitary transformation on $\generichilbertspace$ is an $\innerproduct$-preserving invertible linear map $U$,
\begin{equation}
    \innerproduct[U(a)][U(b)]=\innerproduct[a][b],
\end{equation}
for all $a,b\in\generichilbertspace$. When $\generichilbertspace$ has finite-dimension, we denote the group of unitary transformations by $\unitarygroup{n}$. Its corresponding Lie algebra $\unitaryliealgebra{n}$ is the real Lie algebra of anti-Hermitian linear maps $T$ under commutator,
\begin{equation}
    \innerproduct[T(a)][b]=-\innerproduct[a][T(b)]
\end{equation}
for all $a,b\in\generichilbertspace$.

\section{Abstract Unitary Generators}\label{sec:gen}

Using the inner product $\innerproduct$, we may directly characterize all anti-Hermitian maps in the following way. First, we define a map $\unitarygeneratordeclaration$ by, $\forall a,b,c\in\generichilbertspace$,
\begin{equation}
    \unitarygenerator[a][b] = \mapdefinition{c}{\innerproduct[a][c]b-\innerproduct[b][c]a}.
\end{equation}
We see immediately that $\unitarygenerator[a][b]$ is a $\complexnumbers$-linear map on $\generichilbertspace$, and that $\unitarygenerator$ is $\reals$-bilinear and antisymmetric. We may also easily verify that $\unitarygenerator[a][b]$ is anti-Hermitian in the sense of \eqref{eqn:hermitian-adjoint}. Next, notice that $\unitarygenerator$ fails to be $\complexnumbers$-linear, or $\complexconjugatenumbers$-linear, in either argument; instead, $\unitarygenerator$ is ``middle $\complexconjugatenumbers$-linear'', $\forall a,b\in\generichilbertspace,\alpha\in\complexnumbers$,
\begin{equation}\label{eqn:middle-linearity-unitary-generator}
    \unitarygenerator[\alpha a][b]=\unitarygenerator[a][\complexconjugate{\alpha}b].
\end{equation} 
This means for an $n$-dimensional Hilbert space, there are exactly $n^{2}$ linearly-independent operators of the form $\unitarygenerator[a][b]$. Finally, we calculate that the $\set{\unitarygenerator[a][b]}$ are closed under commutator,
\begin{equation}
    \commutator{\unitarygenerator[a][b]}{\unitarygenerator[c][d]}=\unitarygenerator[\unitarygenerator[a][b][c]][d]+\unitarygenerator[c][\unitarygenerator[a][b][d]].
\end{equation}
These facts together place the real Lie algebra of $\unitarygenerator$ maps under commutator in bijection with $\unitaryliealgebra{n}$. Thus, in exploring the properties of the $\unitarygenerator$ maps, we are directly accessing the properties of the generators of unitary symmetry transformations on $\generichilbertspace$. Furthermore, as we are working without reference to any particular Hilbert space, out results will remain valid for all Hilbert spaces.

\section{Properties of Unitary Generators via Minimal Polynomials}\label{sec:min-pol}

One may imagine that a basis is required to begin to access the properties of the unitary generators, but this is not the case: the most important properties of these linear maps, or any other, may be understood and directly utilized from the operator's minimal polynomial (if one exists). The minimal polynomial $m(x)$ of an operator $A$ is the polynomial of least order such that \cite{halmos_finite-dimensional_1974},
\begin{equation}
    m(A)=0.
\end{equation}
In our case, a short calculation reveals that $\forall a,b\in\generichilbertspace$,
\begin{equation}\label{eqn:t-annihilating}
    \composition{\unitarygenerator[a][b]}{\bigl(\composition{\unitarygenerator[a][b]}{\unitarygenerator[a][b]}+2\unitimaginary\symplecticform\unitarygenerator[a][b]+\biginnerproduct^{2}\id\bigr)}=0,
\end{equation}
where we introduce the following real quantities,
\begin{subequations}
\begin{gather}
    \metric=\half\bigl(\innerproduct[b][a]+\innerproduct[a][b]\bigr)\\
    \symplecticform=\frac{1}{2\unitimaginary}\bigl(\innerproduct[b][a]-\innerproduct[a][b]\bigr)\\
    \biginnerproduct^{2} = \innerproduct[a][a]\innerproduct[b][b]-\innerproduct[a][b]\innerproduct[b][a],\\
    \bigmetric^{2}=\biginnerproduct^{2}+\symplecticform^{2},
\end{gather}
\end{subequations}
for notational ease. It is worth commenting on the existence of \eqref{eqn:t-annihilating}, as in a general Hilbert space it is not clear that such an object should exist, and indeed do not for general operators. In our case, regardless of the dimension of the embedding space, the space considered by acting on $c\in\generichilbertspace$ with $\unitarygenerator[a][b]$ is fundamentally three-dimensional and spanned by $\set{a,b,c}$. After the initial application, which removes any vector components in $c$ orthogonal to the plane spanned by $\set{a,b}$, our action remains confined to this plane. Any operator acting on a finite-dimensional vector space must have a minimal polynomial \cite{axler_linear_2014}, and thus a minimal polynomial must exist for $\unitarygenerator[a][b]$ with at most polynomial order three.

The existence of equation \eqref{eqn:t-annihilating} ensures that a minimal polynomial exists for all $\unitarygenerator[a][b]$, and provides an upper bound for the order of this polynomial. By the Cauchy-Schwartz inequality and the antisymmetry of $\unitarygenerator$, for $\unitarygenerator[a][b]\neq0$ we may choose $\bigmetric>0$, $\biginnerproduct\geq0$, and find that not both of $\symplecticform$ and $\biginnerproduct$ are zero. This means we must consider two, but only two, cases before we proceed further:

\textbf{i)} $\biginnerproduct\neq0$: In this case, 
\begin{equation}\label{eqn:general-t-min-pol}
    m(x)=x(x^{2}+2\unitimaginary\symplecticform x+\biginnerproduct^{2}),
\end{equation}
is minimal for all $\symplecticform\in\reals$. The eigenvalues of this cubic are,
\begin{equation}
    \set{\lambda_{0}=0,\lambda_{1}=-\unitimaginary(\bigmetric+\symplecticform),\lambda_{2}=\unitimaginary(\bigmetric-\symplecticform)},
\end{equation}
and we may use these to algebraically construct projection operators into the corresponding eigenspaces,
\begin{subequations}
\begin{gather}\label{eqn:general-t-proj-first}
    \Pi_{0}=\frac{\bigl(\unitarygenerator[a][b]-\lambda_{1}\id\bigr)\circ\bigl(\unitarygenerator[a][b]-\lambda_{2}\id\bigr)}{\lambda_{1}\lambda_{2}}\\
    \Pi_{1}=\frac{\unitarygenerator[a][b]\circ\bigl(\unitarygenerator[a][b]-\lambda_{2}\id\bigr)}{\lambda_{1}\bigl(\lambda_{1}-\lambda_{2}\bigr)}\\
    \Pi_{2}=\frac{\unitarygenerator[a][b]\circ\bigl(\unitarygenerator[a][b]-\lambda_{1}\id\bigr)}{\lambda_{2}\bigl(\lambda_{2}-\lambda_{1}\bigr)}.\label{eqn:general-t-proj-last}
\end{gather}
\end{subequations}
Note: we have been able to do this without needing to find the corresponding eigenvectors at all! Please refer to \cite{bradshaw_foundations_2024} for a fuller account of this process. For our purposes, we need only know that these projectors resolve the identity,
\begin{equation}\label{eqn:general-t-resolution}
    \id=\Pi_{0}+\Pi_{1}+\Pi_{2}.
\end{equation}

\textbf{ii)} $\biginnerproduct=0$ but $\unitarygenerator[a][b]\neq0$: The existence of this case is subtle and worthy of comment. By the Cauchy-Schwartz inequality, $\biginnerproduct=0$ iff $b=\alpha a$ for some $\alpha\in\complexnumbers$, but the antisymmetry and middle $\complexconjugatenumbers$-linearity of $\unitarygenerator[a][b]$ ensure that $\unitarygenerator[a][b]=0$ iff $b=ka$ for some $k\in\reals$. Thus, this class of $\set{\unitarygenerator[a][b]}$ exist because  ``linear dependence'' is relative to the field of scalars with respect to which an expression is naturally homogenous, and this might differ even for two expressions \textit{on the same vector space}, as we have just observed.

In this case, \eqref{eqn:general-t-min-pol} is not minimal, but,
\begin{equation}
    n(x)=x(x+2\unitimaginary\symplecticform x),
\end{equation}
is. Accordingly, we derive two projection operators into the two eigenspaces,
\begin{subequations}
\begin{gather}\label{eqn:edge-t-proj-first}
    \Pi'_{0}=\frac{\unitarygenerator[a][b]+2\unitimaginary\symplecticform\id}{2\unitimaginary\symplecticform}\\
    \Pi'_{1}=\frac{\unitarygenerator[a][b]}{-2\unitimaginary\symplecticform}.\label{eqn:edge-t-proj-last}
\end{gather}
\end{subequations}
Again, these projectors resolve the identity,
\begin{equation}\label{eqn:edge-t-resolution}
    \id=\Pi'_{0}+\Pi'_{1}.
\end{equation}

\section{Closed-Form Singly-Generated Exponentials and State Mapping}\label{sec:exp}

An arbitrary unitary generator is a real linear combination of $\unitarygenerator[a][b]$, and an arbitrary unitary map is a product of exponentials of such general generators. For our purposes, that is transforming between pairs of arbitrary pure states, we need only a single exponential of a single (carefully chosen) real multiple of $\unitarygenerator[a][b]$. We may use the identity resolutions for our two cases, to immediately find such an exponential in closed-form. In both cases, these are naturally unitary maps, and we have chosen ``normalisations'' on each exponent to render the resulting expressions more simply.

\textbf{i)} $\biginnerproduct\neq0$: Applying $\exp\left(\frac{\theta\unitarygenerator[a][b]}{\bigmetric}\right)$ to both sides of equation \eqref{eqn:general-t-resolution}, and recalling that \eqref{eqn:general-t-proj-first}-\eqref{eqn:general-t-proj-last} are projectors into eigenspaces, we find,
\begin{widetext}
\begin{equation}\label{eqn:general-t-exp}
\begin{aligned}
    \exp\left(\frac{\theta\unitarygenerator[a][b]}{\bigmetric}\right)=\id&+\frac{\unitarygenerator[a][b]}{\biginnerproduct^{2}}\left\{2\unitimaginary\symplecticform+\exp\left(-\sympovermetric[\unitimaginary\theta]\right)\left[\left(\bigmetric+\frac{\symplecticform^{2}}{\bigmetric}\right)\sin(\theta)-2\unitimaginary\symplecticform\cos(\theta)\right]\right\}\\
    &+\frac{\unitarygenerator[a][b]\circ\unitarygenerator[a][b]}{\biginnerproduct^{2}}\left\{1-\exp\left(\sympovermetric[-\unitimaginary\theta]\right)\left[\cos(\theta)+\sympovermetric[\unitimaginary]\sin(\theta)\right]\right\}.
\end{aligned}
\end{equation}
\end{widetext}
It is worth noting that when $\symplecticform=0$, we have $\biginnerproduct=\bigmetric$ and equation \eqref{eqn:general-t-exp} reduces to,
\begin{align}
    \exp\left(\frac{\theta\unitarygenerator[a][b]}{\bigmetric}\right)=\id &+\frac{\unitarygenerator[a][b]}{\bigmetric}\{\sin(\theta)\}\\
    &+\frac{\unitarygenerator[a][b]\circ\unitarygenerator[a][b]}{\bigmetric^{2}}\{1-\cos(\theta)\},
\end{align}
which is the famous ``Rodrigues' rotation formula''. Thus, only for state pairs with $\symplecticform=0$ can we truly call the unitary map they generate a true ``rotation'': without this condition, interference from the phase of $\innerproduct[a][b]$ will alter the action of the map. This is a key qualitative difference between unitary and orthogonal transformations, and by extension between complex Hilbert and real Euclidean spaces.

We may use equation \eqref{eqn:general-t-exp} to compute the angles $\theta$ for which,
\begin{equation}
    \exp\left(\frac{\theta\unitarygenerator[a][b]}{\bigmetric}\right)(a)\propto b,
\end{equation}
and determine the constants of proportionality. Thus, we find,
\begin{equation}
    \exp\left(\frac{\theta'\unitarygenerator[a][b]}{\bigmetric}\right)(a)=\sqrt{\frac{\innerproduct[a][a]}{\innerproduct[b][b]}}\exp\left(\sympovermetric[-\unitimaginary\theta']\right)b
\end{equation}
for angles $\theta'$ satisfying,
\begin{equation}\label{eqn:general-t-sol}
    \cot(\theta')=\frac{\metric}{\bigmetric}.
\end{equation}

That this transformation is only up to a complex multiple is expected: by Wigner's theorem \cite{weinberg_quantum_1995}, unitary transformations preserve the magnitudes of transition probability amplitudes, and so cannot change the magnitudes of states; the $\theta'$-dependent phase factor is a further sign that interference between the two states is affecting the action of the map. There are two fundamental solutions to equation \eqref{eqn:general-t-sol} that differ by $\pi$, which is again expected: when transforming with a periodic operator, one may go the short or the long way around. This solution, and the exponential operator, have been numerically verified for randomly generated state pairs across a wide range of Hilbert space dimensions.

\textbf{ii)} $\biginnerproduct=0$ but $\unitarygenerator[a][b]\neq0$: Applying $\exp\left(\frac{\theta\unitarygenerator[a][b]}{\symplecticform}\right)$ to both sides of equation \eqref{eqn:edge-t-resolution}, and recalling that \eqref{eqn:edge-t-proj-first}-\eqref{eqn:edge-t-proj-last} are projectors into eigenspaces and in this case $\symplecticform\neq0$, we find,
\begin{equation}\label{eqn:edge-t-exp}
    \exp\left(\frac{\theta\unitarygenerator[a][b]}{\symplecticform}\right)=\id+\frac{\unitarygenerator[a][b]}{\symplecticform}\{\exp(-\unitimaginary\theta)\sin(\theta)\}.
\end{equation}
We may again use this expression to solve,
\begin{equation}
    \exp\left(\frac{\theta\unitarygenerator[a][b]}{\symplecticform}\right)(a)\propto b,
\end{equation}
for which we find,
\begin{equation}
    \exp\left(\frac{\theta\unitarygenerator[a][b]}{\symplecticform}\right)(a)=\frac{\innerproduct[b][a]}{\innerproduct[b][b]}\exp(-2\unitimaginary\theta)b,
\end{equation}
for any value of $\theta$. This makes sense, as in this case $b=\alpha a$ for some $\alpha\in\complexnumbers$, and a unitary transformation cannot change the magnitude of a state; thus, this operator can only change the phase of $a$. If we require an angle $\theta'$ such that no additional phase is accrued, this may be achieved when,
\begin{equation}
    \begin{cases}
        \displaystyle\tan(2\theta')=\frac{\symplecticform}{\metric} & \metric\neq0 \\[11pt]
        \displaystyle\theta'=\operatorname{sgn}(\symplecticform)\frac{\pi}{4} & g=0
    \end{cases}
\end{equation}
in which case we have,
\begin{equation}
    \exp\left(\frac{\theta'\unitarygenerator[a][b]}{\symplecticform}\right)(a)=\sqrt{\frac{\innerproduct[a][a]}{\innerproduct[b][b]}}b
\end{equation}

\section{Discussion}\label{sec:disc}

Being able to identify the closed-form exponential unitary transformation between any two given quantum states presents a powerful tool for understanding and improving quantum state preparation (a key subroutine of quantum computation)~\cite{Zhang2022QuantumStatePreparation}. Current approaches for $n$-qubit quantum computation involve initializing a system-wide ground state $\ket{\psi_i} = \ket{0}^{\otimes n}$, then using binary tree encoding to manually ``dial in'' the amplitudes of the $2^n$ different potential elements of the desired $n$-qubit superposition state. Even before considering that each of these amplitude diallings needs to be broken down into 1- or 2-qubit gates between nearest-neighbour qubits (via Gray codes), this approach seems suboptimal.

Identifying the form of the exponential (rather than just the unitary) which you need to apply the desired transformation should aid in this, given a rotation seems more easily decomposable into a small number of shufflings and rotations which are implementable via Gray codes \cite{nielsen_quantum_2010} than the complete basis shifts required through the Gram-Schmidt-based approach. Future work will aim to make this conjecture more rigorous.

While this work focuses on finding the exponential giving the exact unitary we want to find, future work will look at weakening this exactness condition - seeing what insight the algebraic methods applied give for identifying and optimizing approximators to a desired operator.

Additional future work will look at extending the method to transformations between mixed states of the same dimension and purity, and from that (either, by partitioning and tracing out parts of the system, or by adding Lindbladian dynamics) identifying the transformation which takes you from your state to a desired state of equal or lower purity.

In this paper, we have shown how to find the single closed exponential-form unitary operator which maps between any two pure quantum states we choose. In the process of doing this, we defined unitary symmetry in a basis and dimension independent way, we captured the properties of unitary generators by using their minimal polynomials to construct projectors, and we found a closed-form expression for the single unitary exponential that maps a state to any other (up to a complex scalar). Finally, we considered some potential uses for this operator and our methods.

This work has reinforced some the relevance of the algebraic properties of the generators of unitaries. Firstly, it has emphasized that unitary symmetries can be understood by their generators, and that using these generators allows us to describe our unitaries in a basis- and dimension-independent way. It is the algebraic properties of these unitary generators which allowed us to construct the algebraic projectors, which then provided us the closed-form expression for exponential of one generator, which formed the basis of our mapping. Therefore, it is the algebraic properties of these generators which allow us to map any pure state of a given dimension onto any other state of the same dimension with a single exponential (up to a complex scaling). This is a sign the quantum information community should take the algebraic properties of the objects it discusses more seriously.

\textit{Acknowledgements---} The authors thank Nicholas Chancellor for useful discussions. PTJB and JRH acknowledge support from a Royal Society Research Grant (RG/R1/251590), and from JRH's EPSRC Quantum Technologies Career Acceleration Fellowship (UKRI1217). JRH also acknowledges support from an EPSRC Mathematical Sciences Small Grant (UKRI3647).

\bibliographystyle{unsrturl}
\bibliography{ref.bib}

@article{Zhang2022QuantumStatePreparation,
  title = {Quantum State Preparation with Optimal Circuit Depth: Implementations and Applications},
  author = {Zhang, Xiao-Ming and Li, Tongyang and Yuan, Xiao},
  journal = {Phys. Rev. Lett.},
  volume = {129},
  issue = {23},
  pages = {230504},
  numpages = {6},
  year = {2022},
  month = {Nov},
  publisher = {American Physical Society},
  doi = {10.1103/PhysRevLett.129.230504}
}

@book{teschl_mathematical_2014,
    address = {Providence, Rhode Island},
    series = {Graduate {Studies} in {Mathematics}},
    title = {Mathematical {Methods} in {Quantum} {Mechanics}: {With} {Applications} to {Schrodinger} {Operators}},
    isbn = {978-1-4704-1704-8},
    shorttitle = {Mathematical {Methods} in {Quantum} {Mechanics}},
    language = {English},
    number = {157},
    publisher = {American Mathematical Society},
    author = {Teschl, Gerald},
    year = {2014},
}

@book{axler_linear_2014,
    address = {Berlin},
    title = {Linear {Algebra} {Done} {Right}},
    isbn = {978-3-319-11080-6},
    language = {en},
    publisher = {Springer},
    author = {Axler, Sheldon},
    year = {2014},
    note = {Google-Books-ID: 5qYxBQAAQBAJ},
}

@book{nielsen_quantum_2010,
    address = {Cambridge},
    edition = {10th Anniversary},
    title = {Quantum {Computation} and {Quantum} {Information}},
    isbn = {978-1-107-00217-3},
    shorttitle = {Quantum {Computation} and {Quantum} {Information}},
    language = {English},
    publisher = {Cambridge University Press},
    author = {Nielsen, Michael A. and Chuang, Isaac L.},
    year = {2010},
}

@phdthesis{bradshaw_foundations_2024,
    type = {Doctoral},
    title = {Foundations for an {Elementary} {Algebraic} {Theory} of {Systems} with {Arbitrary} {Non}-{Relativistic} {Spin}},
    copyright = {open},
    url = {https://discovery.ucl.ac.uk/id/eprint/10197514/},
    language = {eng},
    urldate = {2025-10-15},
    school = {University College London},
    author = {Bradshaw, Peter Thomas Joseph},
    month = sep,
    year = {2024},
}

@book{halmos_finite-dimensional_1974,
    address = {New York, NY},
    series = {Undergraduate {Texts} in {Mathematics}},
    title = {Finite-{Dimensional} {Vector} {Spaces}},
    isbn = {978-1-4612-6389-0 978-1-4612-6387-6},
    url = {http://link.springer.com/10.1007/978-1-4612-6387-6},
    urldate = {2023-07-13},
    publisher = {Springer},
    author = {Halmos, Paul R.},
    editor = {Axler, S. and Gehring, F. W. and Ribet, K. A.},
    year = {1974},
    doi = {10.1007/978-1-4612-6387-6},
}

@book{weinberg_quantum_1995,
    address = {Cambridge},
    title = {The {Quantum} {Theory} of {Fields}: {Volume} 1: {Foundations}},
    volume = {1},
    isbn = {978-0-521-67053-1},
    shorttitle = {The {Quantum} {Theory} of {Fields}},
    url = {https://www.cambridge.org/core/books/quantum-theory-of-fields/22986119910BF6A2EFE42684801A3BDF},
    urldate = {2023-05-27},
    publisher = {Cambridge University Press},
    author = {Weinberg, Steven},
    year = {1995},
    doi = {10.1017/CBO9781139644167},
}

\end{document}